# Efficient real-time spin readout of nitrogen-vacancy centers based on Bayesian estimation

Jixing Zhang, [1†\*] Tianzheng Liu, [2, 5†] Sigang Xia, [3] Guodong Bian, [2] Pengcheng Fan, [2] Mingxin Li [2], Sixian Wang, [2] Xiangyun Li, [2] Chen Zhang, [1] Shaoda Zhang, [4] and Heng Yuan, [2\*]

[1] 3rd Institute of Physics, University of Stuttgart, Stuttgart 70569, Germany;

[2] School of Instrumentation and Optoelectronic Engineering, Beihang University, Beijing 1000191, China;

[3] Shanghai Electro-Mechanical Engineering Institute, Shanghai 201109, China;

[4] Shenzhen Cofoe Biotechnology Co., Ltd, Shenzhen 518000, China.

[5] Goertek Inc., Weifang 261031, China.

\*Corresponding author: jixing.zhang@pi3.uni-stuttgart.de; hengyuan@buaa.edu.cn

†These authors contributed equally to this work.

*Abstract-* **In this work, to improve the spin readout efficiency of the nitrogen vacancy (NV) center, a real-time Bayesian estimation algorithm is proposed, which combines both the prior probability distribution and the fluorescence likelihood function established by the implementation of the NV center dynamics model. The theoretical surpass of the Cramer-Rao lower bound of the readout variance and the improvement of the readout efficiency in the simulation indicate that our approach is an appealing alternative to the conventional photon summation method. The Bayesian real-time estimation readout was experimentally realized by combining a high-performance acquisition and processing hardware, and the Rabi oscillation experiments divulged that the signal-to-noise ratio of our approach was improved by 28.6%. Therefore, it is anticipated that the employed Bayesian estimation readout will effectively present superior sensing capabilities of the NV ensemble, and foster the further development of compact and scalable quantum sensors and consequently novel quantum information processing devices on a monolithic platform.**

## 1. Introduction

As an extraordinary solid spin material, the diamond nitrogen vacancy (NV) center presents an exciting frontier in the high-precision sensing and measurement fields[1][2]. At present, this solid-state quantum system with excellent spin coherence properties has been successfully applied for various applications, including magnetic sensing[3][4][5], inertial sensing[6][7][8], temperature sensing[9][10], quantum information processing[11][12], and biomedical measurements[13][14]. The efficiency of the electron spin projection readout, an indispensable building block for the emerged quantum technologies, directly determines the overall performance of the sensing element[15][16][17]. The conventional readout approach is based on the spin-dependent photoluminescence intensity of the NV center, namely the photon summation (PS) technique, which is, however, seriously affected by the photon shot noise[18][19]. So far, several approaches have been proposed to mitigate the spin readout noise, among which are the spin-to-charge conversion readout method[20] (requiring the employment of long spin readout times) and photoelectric readout procedure[21] (consisting in adding the complementary measurement noise due to the fluctuations in the applied electric field). Moreover, some post-processing methods, such as the machine learning technique[22], require redundant empirical information and cannot perform in real time.

Bayesian estimation is a promising scheme combining empirical prior and sample likelihood. As far as quantum sensing applications are concerned, various Bayesian estimation schemes have recently been proposed to improve the performance of information processing and have already been experimentally implemented[23][24][25]. The diversity of prior selection schemes and concise information processing make up for the subjectivity deficiency of classical information processing and coordinate the randomness and sufficiency of quantum information. Along these lines, a Bayesian estimation method for the spin readout process is proposed, and it is anticipated that our approach can accurately extract the effective spin information of the fluorescence signal of the noisy system and realize the real-time readout technique.

In this work, based on the dynamic model of the diamond NV center, the fluorescence readout process and the respective likelihood function were analytically deduced. By combining the prior information and the likelihood function, a readout method based on the Bayesian estimation algorithm was proposed. This method was compared with the conventional PS method and its influence on the readout efficiency was thoroughly investigated. The Cramer-Rao lower bound of the readout variance was also derived and used as a benchmark to compare the theoretical variance of the conventional PS method and the Bayesian method with classical prior distributions. Then, the readout mean square error (MSE) of the two methods was simulated, and the real-time performance of the Bayesian method was discussed. Finally, based on the system-on-chip platform, a real-time spin readout process was realized in the experiment, and the Rabi oscillation experiments revealed that the signal-to-noise ratio (SNR) of the Bayesian method was improved by 28.6% compared with the conventional PS method.

## 2. Methods
*2.1 NV energy levels and the dynamic model*

The energy level structure of the NV center at room temperature is schematically illustrated in Fig. 1(a)[26][27][28]. The ground state is composed of a spin-triplet state, and this state includes the sub-states of $|m_s = 0\rangle$ (1) and $|m_s = \pm 1\rangle$ (2). The excited state is also a spin-triplet state, which contains the sub-states of $|m_s = 0\rangle$ (3) and $|m_s = \pm 1\rangle$ (4). The metastable state (5) is a single state.

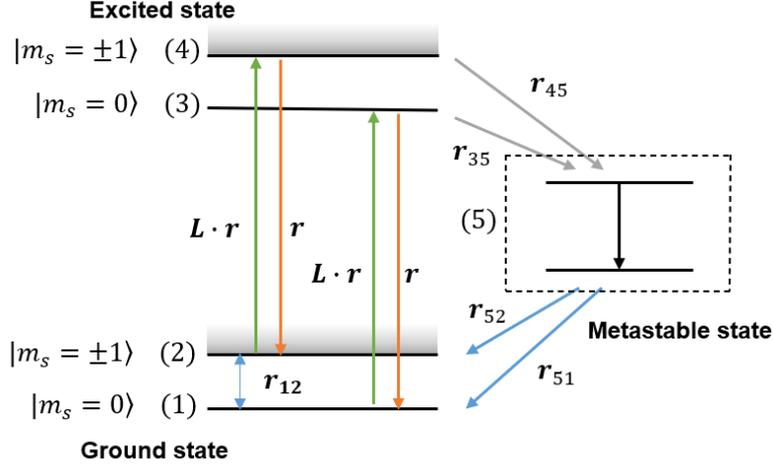

Figure 1. Depiction of the NV energy level structure and the corresponding dynamics.

A 532 nm laser pumps the ground state electrons to the excited state and the pumping rate is $L \cdot r$, where $L$ is the relative pumping rate that is proportional to the optical power density of the laser. Some electrons in the excited state directly decay to the ground state with a relaxation rate $r$ and fluoresce. Others decay to the ground state through metastable state with relaxation rates $r_{35}$, $r_{45}$, $r_{51}$, $r_{52}$ and emit infrared light. The spin relaxation rate is signified by $r_{12}$, which is related to the longitudinal relaxation time $T_1$. The data for the energy level structure are derived from Ref.[29]. The optical rate equation can be derived from the energy level diagram as is shown in Fig. 1[30][31]:

$$\begin{aligned}
\frac{dp_1(t)}{dt} &= r \cdot p_3(t) + r_{51} \cdot p_5(t) - L \cdot r \cdot p_1(t) + r_{12} \cdot (p_2(t) - p_1(t)) \\
\frac{dp_2(t)}{dt} &= r \cdot p_4(t) + r_{52} \cdot p_5(t) - L \cdot r \cdot p_2(t) - r_{12} \cdot (p_2(t) - p_1(t)) \\
\frac{dp_3(t)}{dt} &= L \cdot r \cdot p_1(t) - r \cdot p_3(t) - r_{35} \cdot p_3(t) \\
\frac{dp_4(t)}{dt} &= L \cdot r \cdot p_2(t) - r \cdot p_4(t) - r_{45} \cdot p_4(t) \\
\frac{dp_5(t)}{dt} &= r_{35} \cdot p_3(t) + r_{45} \cdot p_4(t) - r_{51} \cdot p_5(t) - r_{52} \cdot p_5(t)
\end{aligned} \quad (1)$$

where $p_i(t)$ is denoted as the $i$ state's probability distribution. When the system without the laser excitation, electrons are located in the ground state. Set the population of $m_s = 0$ in the ground state is $\rho$, and the rate equation can be further reduced to $\vec{p}(t) = e^{Mt} \cdot \vec{p}(0)$, where $\vec{p}(t) = (p_1(t), p_2(t), p_3(t), p_4(t), p_5(t))^T$, $\vec{p}(0) = (\rho, 1-\rho, 0,0,0)^T$ and

$$M = \begin{bmatrix} -L \cdot r - r_{12} & r_{12} & r & 0 & r_{51} \\ r_{12} & -L \cdot r - r_{12} & 0 & r & r_{52} \\ L \cdot r & 0 & -r - r_{35} & 0 & 0 \\ 0 & L \cdot r & 0 & -r - r_{45} & 0 \\ 0 & 0 & r_{35} & r_{45} & -r_{51} - r_{52} \end{bmatrix}. \quad (2)$$

During the readout process, the fluorescence produced by the NV center is proportional to the excited state population $p_3(t) + p_4(t)$[32][33]. If the fluorescence signal is collected within the time interval $\Delta t = t_n - t_{n-1}$, the function $f_n$ can expressed as follows:

$$\langle f_n \rangle = \lambda \cdot [0,0,1,1,0] \cdot \vec{p}(t_n) \cdot \Delta t, \quad (3)$$

where $\lambda$ represents the collection efficiency and the number of NV centers. By simplifying Eq. (3), the theoretical fluorescence signal determined by $\rho$ at time $t_n$ can be obtained by the following equation:

$$\langle f_n(\rho, t_n)\rangle = \lambda \cdot [0,0,1,1,0] \cdot e^{Mt_n} \cdot (\rho, 1-\rho, 0,0,0)^T \cdot \Delta t = A(t_n) \cdot \rho + B(t_n), \tag{4}$$

where A (B) is the dependent (independent) term of $\rho$. In practice, the photon shot noise that satisfies the Poisson distribution will be generated during the collection process, and the likelihood function for the fluorescence signal $f_n$ at time $t_n$ is defined as follows[34][35]:

$$P(f_n|\rho) = \frac{[\langle f_n(\rho, t_n)\rangle]^{f_n}}{f_n!} \cdot \exp[-\langle f_n(\rho, t_n)\rangle]. \tag{5}$$

The utilization efficiency of the information contained in noisy fluorescence signal determines the readout accuracy[36][37][38], and quantization of noise paving the way for optimizing the readout process.

*2.2 Bayesian estimation spin readout method*

The fluorescence signal with noise can be simulated based on Eq. (5), as is illustrated in Fig. 2(a). The conventional PS method simply sums up the fluorescence signal, which is severely hindered by the photon shot noise and technical noise[39][40]. Here, based on the parameter estimation in mathematical statistics[41][42][43], a Bayesian estimation algorithm is proposed to fruitfully exploit the spin readout information. The applied Bayesian estimation protocol is displayed in the Fig. 2(b). Combining the likelihood function $P(f_n|\rho)$ of the fluorescence $f_n$ with the previous probability distribution of $P(\rho|f_{n-1}^{meas})$, the probability distribution $P(\rho|f_n^{meas})$ can be updated as follows:

$$P(\rho|f_n^{meas}) = \frac{P(f_n|\rho) \cdot P(\rho|f_n^{meas})}{P(f_n)}, \tag{6}$$

where $f_n^{meas} = (f_1, f_2, \cdots, f_n)$ represents the set of measured fluorescence, and $P(f_n)$ is the normalized constant. An example of the working principles of the Bayesian estimation protocol is presented in Fig. 2(c), where the prior distribution $P(\rho|f_0^{meas})$ is uniform. In order to improve the efficiency of estimation, the 90% confidence interval of probability distribution $P(\rho|f_n^{meas})$ is used as a measure of $P(\rho|f_n^{meas})$ variance:

$$P(c_1 \leq \rho^E \leq c_2) = 0.9, \tag{7}$$

where $\rho^E$ is the estimated result, which is the expectation of the probability distribution $P(\rho|f_n^{meas})$, $c_1$ and $c_2$ are the lower limit and upper limit of confidence interval respectively.

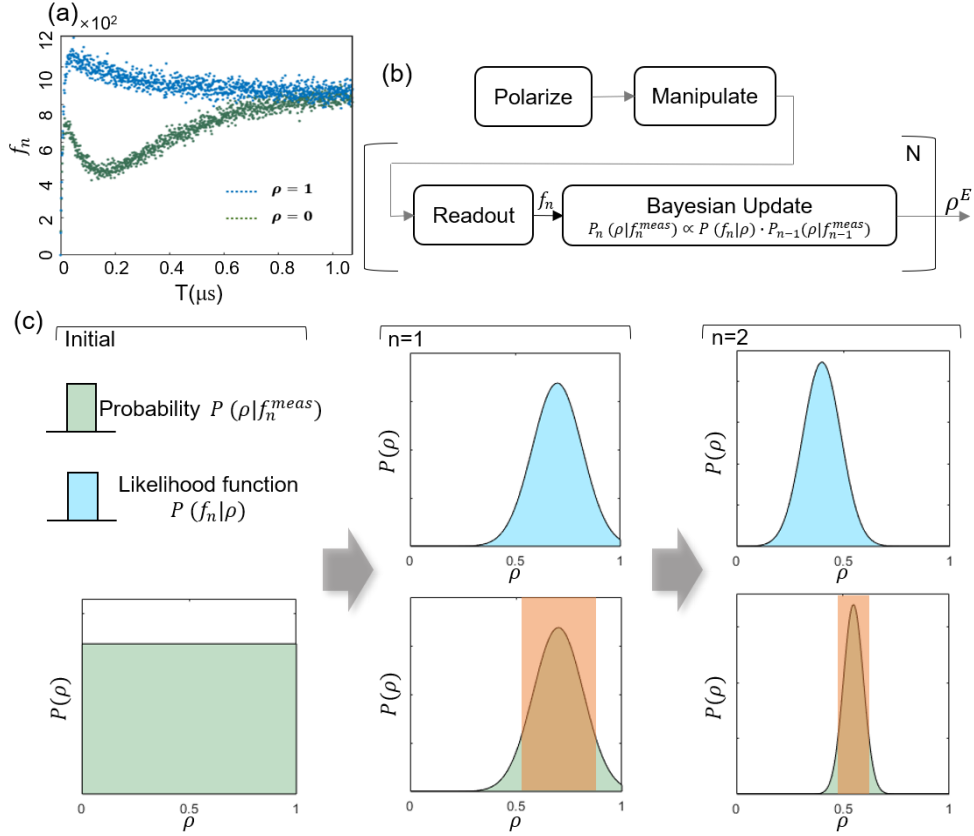

Figure 2. (a) Distribution of the readout fluorescence signal with the photon shot noise. The figure shows the most ($\rho = 1$) and least ($\rho = 0$) fluorescence at readout. The readout results show the most ($\rho = 1$) and least ($\rho = 0$) fluorescence. (b) The Bayesian estimation protocols. The Bayesian measurement estimation protocol consists of a sequence of the readout and update operations after the polarization and manipulation. The fluorescence signal was used to update the estimate result $\rho^E$ after each readout process. (c) An example of a Bayesian update process. At step $n$, the probability distribution $P(\rho|f_n^{meas})$ is calculated by combining the likelihood function $P(f_n|\rho)$ of the fluorescence $f_n$ and the previous probability distribution $P(\rho|f_{n-1}^{meas})$. The orange shaded regions represents the 90% confidence interval for the corresponding probability distribution $P(\rho|f_n^{meas})$.

The readout accuracy can be evaluated by estimating the variance of the signal estimation. It is well-known that the lower bound of the estimated variance, namely the Cramer-Rao lower bound (CRLB), can be reached by obtaining all the information of the signal, which is defined as the reciprocal of Fisher information $I(\rho)$ [44]. The fluorescence signal $f_n$ satisfies the independent Poisson distribution, and the CRLB can be derived as follows:

$$\sigma_{CRLB}^2 = 1/I(\rho) = 1/\sum_{n=1}^{N}\{A(t_n)^2/[A(t_n)\cdot\rho + B(t_n)]\}. \tag{8}$$

The readout variance of the conventional PS method can be given by Eq. (4):

$$\sigma_{PS}^2 = 1/\{\sum_{n=1}^{N}A(t_n)^2 /[\sum_{n=1}^{N}A(t_n)\cdot\rho + B(t_n)]\}. \tag{9}$$

When $N > 1$, $\sigma_{PS}^2 > \sigma_{CRLB}^2$, the conventional PS method cannot reach the lower bound of the estimated variance.

The readout variance of the Bayesian method can be derived from the mathematical statistics theories. When the update number $N$ is determined, the estimation result $\rho^E$ can be expressed as follows:

$$\rho^E = \arg\max P(\rho|f_N^{meas}) = \arg\max\left(\frac{\prod_{n=1}^N P(f_n|\rho)}{\prod_{n=1}^N P(f_n)} \cdot P(\rho|f_0^{meas})\right). \quad (10)$$

$\rho^E$ can be solved by the logarithmic probability function $P(\rho|f_N^{meas})$:

$$\ln\left(P(\rho|f_N^{meas})\right) = \sum_{n=1}^N f_n \cdot \ln(A(t_n) \cdot \rho + B(t_n)) - \sum_{n=1}^N A(t_n) \cdot \rho + B(t_n) + \ln P(\rho|f_0^{meas}) + \text{const.} \quad (11)$$

Let $\frac{\partial}{\partial \rho}\ln(P(\rho|f_N^{meas})) = 0$ to obtain $\rho^E$:

$$\sum_{n=1}^N f_n \cdot \frac{A(t_n)}{(A(t_n)\cdot\rho^E + B(t_n))} - \sum_{n=1}^N A(t_n) + \frac{\partial}{\partial \rho}\ln P(\rho^E|f_0^{meas}) = 0. \quad (12)$$

This formula cannot be solved mathematically, but it can be approximated to a Gaussian distribution when the number of the samples that satisfy Poisson distribution is sufficient. By replacing $f_n \sim N(\langle f_n(\rho, t_n)\rangle, \langle f_n(\rho, t_n)\rangle))$ to Eq. (12), which is simplified by the logarithmic processing and setting the partial derivative to 0:

$$-\sum_{n=1}^N \frac{A(t_n)}{(A(t_n)\cdot\rho + B(t_n))}(A(t_n)\cdot\rho^E + B(t_n) - f_n) + \frac{\partial}{\partial \rho}\ln P(\rho^E|f_0^{meas}) = 0. \quad (13)$$

According to the Eq. (13), the selection of prior distribution $P(\rho^E|f_0^{meas})$ has a great influence on the estimation result. The calculation process of the estimated variance of the three typical prior distributions is analytically presented in Appendix A[45]. When the priors are flat and Jeffreys, the estimated variance can be estimated by the following equation:

$$\sigma_{flat}^2 = \sigma_{Jeffreys}^2 = 1/\frac{1}{I(\rho)}. \quad (14)$$

It can be seen that the accuracy of Bayesian estimation of these two priors is the same to that of the maximum likelihood estimation, a common post-processing optimization method, and both accuracies can reach the CRLB. Notably, when the prior the Gaussian distribution $N(\rho_0, \sigma_0^2)$ is selected for the prior, the estimated variance can exceed CRLB:

$$\sigma_{conjugate}^2 = \frac{1}{I(\rho) + \frac{2}{\sigma_0^2} + \frac{1}{\sigma_0^4 \cdot I(\rho)}}. \quad (15)$$

This effect can be interpreted by considering that CRLB is the lower bound of the estimated variance obtained from all the information of the fluorescence signal. However, if prior as the empirical information outside the region of the signal provides reliable subjective information, the estimated variance can be lower than CRLB. The exceeding degree can be also intuitively compared by CRLB:

$$\sigma_{CRLB}^2 - \sigma_{conjugate}^2 = \left(1 - \frac{1}{(k+1)^2}\right) \cdot \sigma_{CRLB}^2, \quad (16)$$

where $k = \sigma_{CRLB}^2/\sigma_0^2$ represents the ratio of the variance of Gaussian prior distribution to CRLB. As it can be ascertained from Eq. (16), the reduction of the variance of the prior distribution $\sigma_0^2$ can improve the exceeding degree, but a decrease in prior variance of prior will make the selection of expectation $\rho_0$ extremely difficult. As far as the selection of the prior distribution is concerned, it will be further elaborated in the experimental verification part.

*2.3 The simulation evaluation*

The following section deals with the simulation analysis based on the Bayesian method as the noise suppression technique. By using the fluorescence simulation with a Poisson noise as input, the readout variance can be measured by employing the mean square error (MSE) of the unbiased estimation as follows:

$$\text{MSE} = \langle(\rho^E - \rho)^2\rangle. \quad (17)$$

For real-time requirements, the performance of the hardware is one of the important factors in determining the readout efficiency. Under this perspective, Fig. 3(a) displays the simulated influence of the fluorescence sampling time $dt$ on MSE during the signal collection. In general, the sampling time dt has a lower limit in range of 1 ns because the fluorescence detector used in ensemble situation, such as photodiode, has a bandwidth smaller than 1 GHz. However, in practice, to ensure the real-time performance, a certain amount of time needs to be reserved for updating the calculation after each fluorescence acquisition, so the same resolution may not be achieved under applying the same experimental conditions. As is shown in the dotted line and the shadow above in Fig. 3(a), when the resolution of the conventional PS method is 1 ns, the Bayesian method provides better readout effect when the resolution is less than about 20 ns.

In addition, during the process of updating after the fluorescence sampling, both the probability distribution $P(\rho|f_n^{meas})$ and likelihood function $P(f_n|\rho)$ can be conveniently utilized for the matrix operation in hardware through the discretization of the same sampling points. The MSE at different sampling points is illustrated in Fig. 3(b). The sampling points can be considered as a linear correlation effect of the operation time of an update. In practice, it is necessary to choose an appropriate discrete sampling rate to balance the readout accuracy and measurement operation time to improve the readout efficiency.

The following section examines the performance comparison between the Bayesian and the conventional PS methods. Fig. 3(c) depicts the MSE as a function of readout time for both methods. It is evident that the MSE in the conventional PS method enriches an optimal value as the time increases, which is attributed to the accumulation of the shot noise induced by the photon summation, while the MSE in the Bayesian method decreases with time. It is known that the high fluorescence intensity will suppress the noise relatively[46]. In this respect, the MSE plotted versus the relative pumping rate $L$ is shown in Fig. 3(d). It can be seen that both methods provide more accurate results as the value $L$ increases, but the Bayesian method can better suppress the noise by using a greater $L$.

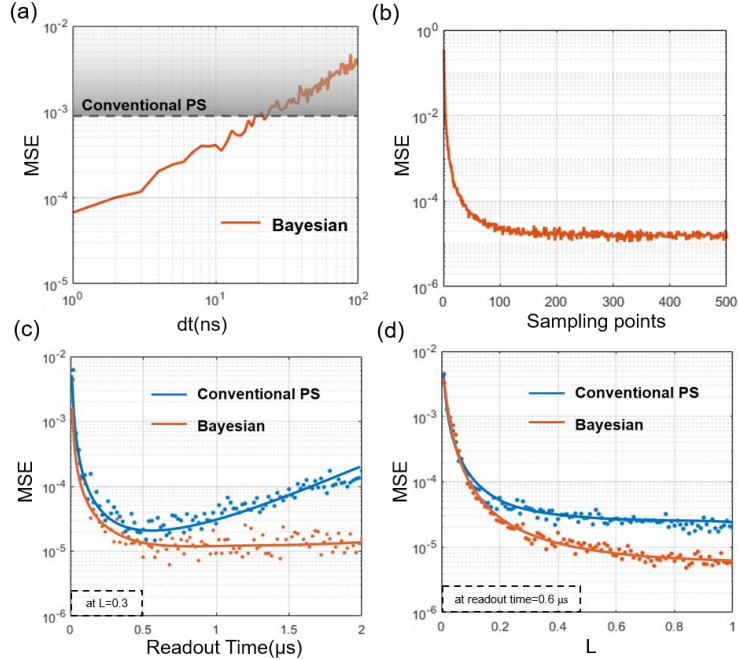

Figure 3. (a) Depiction of the MSE in the Bayesian method at different fluorescence sampling resolutions $dt$. The dashed line highlights the MSE of the conventional PS method, which is about 0.9*10$^{-3}$. (b) The MSE of the Bayesian method under applying different sampling points. (c) The MSE in the conventional PS and the Bayesian method at different readout times (@ L = 0.3). Each point is the mean value of 100 test samples, and the curve is the fitting of test samples. The optimal time of the conventional PS method is 0.6 μs. (d) The MSE in the conventional PS and the Bayesian method at different relative pumping rates $L$. The readout time of the two methods is set as 0.6 μs, which is the optimal value of the conventional PS method and within the optimal range of the Bayesian method. Each point represents the mean value of 100 test samples, and the curve is the fitting of test samples.

## 3. Experimental verification

*3.1 Experimental setup*

In this part, the theoretical method is verified and discussed experimentally. A homemade confocal experimental system was used for the NV spin readout, as is shown in Fig. 4(a). The laser (MSL-FN-532 nm) had a maximum power of 500 mW, and its power stability was 1%. The optical power density of the optical path was controlled by employing a λ/2 plate and the polarization beam splitter (PBS). The acoustic optical modulator (AOM, Gooch&Housego 3200-1911), controlled by a pulse generator (Spincore), was used to generate pulses for both the polarization and readout processes. It was composed of two lenses of the same focal length on either side to focus the laser beam. An objective lens with a 50×/0.55 numerical aperture and a working distance of 7.9 mm was used to focus the laser beam onto a diamond sample. The microwave source (Keysight N5181B) was controlled by a microwave switch (Minicircuits ZASW-2-50dr +), which was also controlled by the pulse generator, connected to the gold wire on the printed circuit board (PCB) to manipulate the CVD-based sample with N impurity concentration ≈ 50 ppm. The sample was also treated under $5 \times 10^{17}$ ea cm$^{-2}$ with 10 MeV electron irradiation and 4 h annealing at 800 ℃. On top of that, the home-made Helmholtz coils produced a static magnetic field $B$ = 30 Gs along the diamond NV [1,1,1] axis. The microwave frequency $f$ was set to the value of 2.956 MHz, which was obtained by a single peak of the optically detected magnetic resonance spectrum. The microwave source power was set to -2 dBm, which was powered up by a +9 dBm type power amplifier (Mini-Circuits ZHL-16W-43+). An avalanche photodiode (APD, Thorlab APD120A/M) with a steady-state voltage of about 0.5 V and root-mean-square error less than 0.015 %, was also utilized for the fluorescence collector. The APD outputs to an analog-to-digital converter (ADC, ADI AD9680-1250) for sampling. A system-on-chip (ZYNQ Xilinx ZCU102), which combines field-programmable gate array (FPGA) and advanced reduced instruction-set computer machine (ARM), was used for the real-time readout update calculation, and the specific architecture of the system-on-chip is shown in Appendix B. The protocol and fluorescence signals are shown in Fig. 4(b).

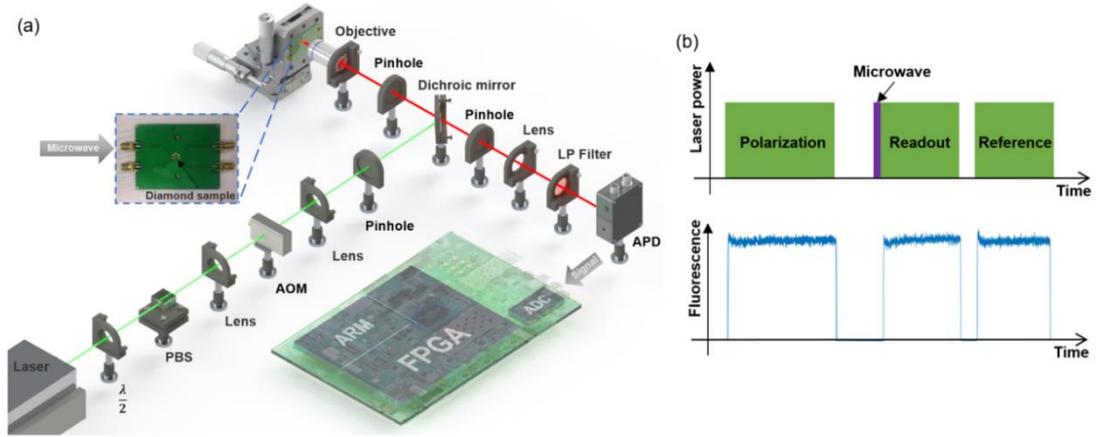

Figure 4. (a) Depiction of the three-dimensional diagram of the experimental setup. (b) Schematic illustration of the experimental sequence. The upper part is the sequence used in the experiment, where the green is the laser pulse and the purple is the microwave pulse. The reference pulse is to ensure that the singlet completely relaxes to the ground state. The polarization (8 μs) and readout (5 μs) pulses were spaced 3 μs apart, whereas the readout and reference (5 μs) pulse were spaced 2 μs apart. The lower part is the fluorescence generated by diamond samples under this sequence.

*3.2 Experimental results*

In Bayesian updating, both the coefficients $\lambda$ and $L$ in the likelihood function need to be determined, and the discriminant method is presented in Appendix C. Subsequently, the readout variance and update time under different discrete sampling points were also verified as was revealed in Fig. 5(a). The increase of the sampling points can significantly reduce the readout variance when the sampling points are smaller than 400, but the application of excessive sampling points has a smaller impact on the readout and a larger impact on the calculation time of update, leading to the deterioration of the real-time performance. The Bayesian and the conventional PS methods were used for the repeated readout, and the variance $D^2$ were compared, as is shown in Fig. 5(b). Since the repeated process could not clearly obtain the prior information, the prior was set as flat. It can be observed that the experimental results are fundamentally identical to the simulations in Fig. 3(c). Under the optimal time, the noise suppression ability of the Bayesian method is improved by about 40.9% compared with the conventional PS method.

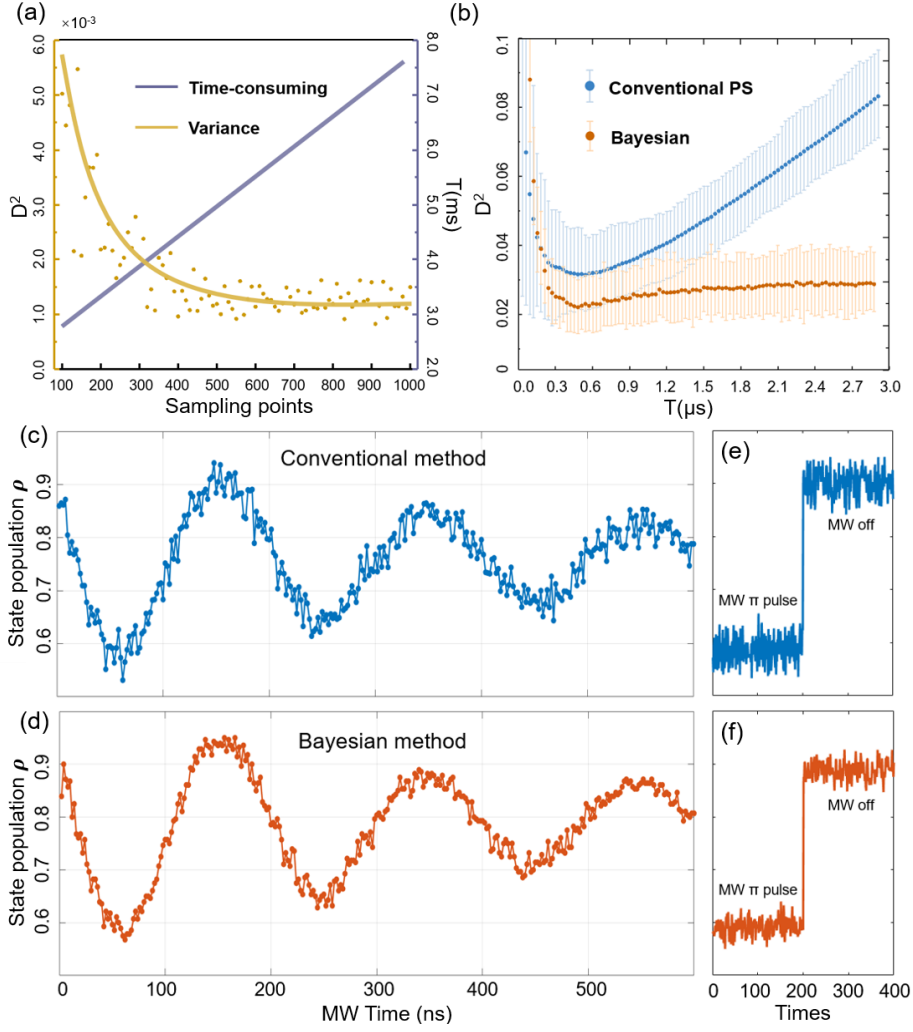

Figure 5. (a) The readout $D^2$ and the Bayesian update are time-consuming under the application of different discrete sampling points. Each point in the figure is the mean error from the average over 100 measurements, and the readout time is set at 0.5 μs. The time-consuming was measured by the clock in the hardware and is linear with discrete sampling points. (b) Readout $D^2$ of the conventional PS and the Bayesian methods at different readout times. The microwave time was set as π pulse. Each point in the figure signifies the mean deviation square from the execution of 100 measurements, which is defined as the variance, and its upper and lower limits correspond to the 95 % confidence intervals. The blue dot represents the conventional PS method, and the optimal time is about 0.5 μs. The red dot stands for the Bayesian method, which exhibits an optimal time of about 0.4 μs. (c) Conventional PS method of the Rabi oscillation raw data. The readout time was set to 0.5 μs. (d) Bayesian method of the Rabi oscillation raw data. The readout time was set to 0.5 μs. (e) 400 readout tests of conventional PS method. The first 200 times were the readout results with microwave (MW) π pulse, and the last 200 times were the readout results without microwave pulse. The SNR was 9.03. (f) 400 readout tests of Bayesian method. The test scheme was the same as Fig. 5(e). The SNR was 11.61.

Finally, the Rabi oscillation results of the conventional PS and the Bayesian methods are compared. Since the Rabi oscillation can be obtained theoretically, the conjugated Gaussian distribution can be used as the prior distribution of the Bayesian estimation, namely $(\rho|f_0^{meas}) \sim N(\rho_0, \sigma_0^2)$. Parameters of prior distribution: 1) The expectation $\rho_0$ under the enforcement of a different microwave time can be estimated by the probability distribution $p_1$ of the $|m_s = 0\rangle$ state in the electron spin mechanics model[17][47]:

$$\begin{aligned}
\frac{dp_{12}}{dt} &= -\frac{p_{12}}{T_2^*} + ip_{12}\left(f - (\gamma_e B + D_{gs})\right) + i\frac{\Omega}{2}(p_1 - p_2) \\
\frac{dp_{21}}{dt} &= -\frac{p_{21}}{T_2^*} + ip_{21}\left(f - (\gamma_e B + D_{gs})\right) + i\frac{\Omega}{2}(p_2 - p_1) \\
\frac{dp_1}{dt} &= i\frac{\Omega}{2}(p_{12} - p_{21}) - \frac{1}{T_1}(p_1 - p_2) \\
\frac{dp_2}{dt} &= i\frac{\Omega}{2}(p_{21} - p_{12}) - \frac{1}{T_1}(p_2 - p_1)
\end{aligned}, \quad (18)$$

where $\Omega$ stands for Rabi frequency, which is equal to the gyromagnetic ratio $\gamma_e$ multiplied by the microwave power $B_{MW}$; $f$ represents the microwave frequency; $p_{ij}$ denotes the non-diagonal term of the density matrix, and $D_{gs}$ is the zero field splitting. The microwave field intensity simulation at the center point of the microwave resonator that was used in the experiment was about 0.178 mT, corresponding to the Rabi frequency $\Omega = 5.11$ MHz. The measured values of the dephasing time $T_2^*$ and the longitudinal relaxation time $T_1$ were 288 ns and 1.03 ms, respectively. 2) The variance $\sigma_0^2$ of the Gaussian prior was set to be 3-5 times of $\sigma_{CRLB}^2$ by considering the measurement error of the dynamic model parameters and the influence of the technical noise.

To ensure the consistency of the comparison, the fluorescence signal was collected by employing an ADC and processed in hardware in parallel with the two methods. The Rabi oscillations of both the conventional PS and Bayesian methods are depicted in Figs. 5(c) and 5(d), where the Bayesian method exhibits a better noise suppression effect. In order to quantitatively compare the readout effects of the two methods, 400 readout tests were repeated as depicted in Figs. 5(e) and (f), of which 200 were microwave π pulse readout and 200 were non-microwave readout. The peak-to-peak value was set as a signal, and the standard deviation was set as noise. The calculation demonstrated that the SNR of the Bayesian method was 28.6% higher than that of the conventional PS method.

## 4. Conclusions

In summary, to improve the spin readout efficiency of the ensemble NV center measurement, an improved readout method based on the Bayesian estimation algorithm was proposed. In this method, the spin state was estimated by combining the prior distribution and the fluorescence likelihood function based on the NV dynamics model. The introduction of the prior information is expected to achieve an experimental breakthrough in the Cramer-Rao lower bound of the variance during the spin readout process. According to the simulation results, the method has higher readout accuracy and better real-time performance. The high-performance system-on-chip was utilized to realize the real-time Bayesian estimation spin readout process, and the extracted outcomes divulge that the readout noise suppression ability was increased by 28.6% than the conventional PS method. Therefore, our method provides an appealing alternative to the conventional approach and presents superior sensing capabilities of the pulsed NV center. Additionally, it provides fruitful insights for the miniaturization and high-precision measurement of solid spin systems.

## Acknowledgment

This work was supported by the National Natural Science Foundation of China under Grant Nos. 62173020 and 62103381, and China Postdoctoral Science Foundation 2021T140624.

**Appendix**

*Appendix A. Estimation variance of the different prior distributions*

The results of the Bayesian estimation can be given by the following formula:

$$-\sum_{n=1}^{N} \frac{A(t_n)}{(A(t_n)\cdot\rho+B(t_n))}(A(t_n)\cdot\rho^E+B(t_n)-f_n)+\frac{\partial}{\partial\rho}\ln P(\rho|f_0^{meas})=0, \quad (A1)$$

where $P(\rho|f_0^{meas})$ represents the prior distribution given in advance. The implementation of different prior distributions such as the flat prior, the Jeffreys prior, and the conjugate prior has a great influence on the estimation results that will be deduced below.

1) The flat prior. Due to the limitation of the fluorescence signal points, the uniform distribution can be used as the flat prior, which is defined as $P(\rho|f_0^{meas})\sim U(0,1)$. The estimated result can be obtained by substituting the distribution into Eq. (A1):

$$\hat{\rho}=\frac{\sum_{n=1}^{N}\frac{A(t_n)}{(A(t_n)\cdot\rho+B(t_n))}(f_n-B(t_n))}{\sum_{n=1}^{N}\frac{A(t_n)^2}{(A(t_n)\cdot\rho+B(t_n))}}. \quad (A2)$$

It can be observed that the variance of the fluorescence $f_n$ satisfying the Poisson distribution is equal to its expectation $\langle f_n(\rho,t_n)\rangle$, so the variance $\sigma_{flat}^2$ of the flat prior can be determined as follows:

$$\sigma_{flat}^2=1/\sum_{n=1}^{N}\frac{A(t_n)^2}{A(t_n)\cdot\rho+B(t_n)}=\frac{1}{I(\rho)}. \quad (A3)$$

2) The Jeffreys prior. This is a no-information prior that uses the posterior information of the sample to take full effect in the estimation [48]. Both the Jeffreys and the flat prior provide as little information as possible so that the estimation result is only furnished by the fluorescence likelihood function, namely Eq. (5). The Jeffreys prior distribution is the determinant of the Fisher information matrix under the square root, which is defined for a single estimated parameter as follows:

$$P(\rho|f_0^{meas})=\sqrt{I(\rho)}, \quad (A4)$$

where $I(\rho)$ is the Fisher information. Substituting Eq. (A4) into Eq. (A1) allows one to get a simplified form of the latter one, that is

$$-\sum_{n=1}^{N}\frac{A(t_n)}{(A(t_n)\cdot\rho+B(t_n))}(A(t_n)\cdot\rho^E+B(t_n)-f_n)-\frac{1}{2}\sum_{n=1}^{N}\frac{A(t_n)}{(A(t_n)\cdot\rho+B(t_n))}=0. \quad (A5)$$

After the simplification process, the estimated results are as follows:

$$\rho^E=\frac{\sum_{n=1}^{N}\frac{A(t_n)}{(A(t_n)\cdot\rho+B(t_n))}\left(f_n-B(t_n)-\frac{1}{2}\right)}{\sum_{n=1}^{N}\frac{A(t_n)^2}{(A(t_n)\cdot\rho+B(t_n))}}. \quad (A6)$$

The variance $\sigma_{Jeffreys}^2$ of the Jeffreys prior can be further obtained by substituting the Poisson variance of the fluorescence signal $f_n$:

$$\sigma_{Jeffreys}^2=\frac{1}{\sum_{n=1}^{N}\frac{A(t_n)^2}{A(t_n)\cdot\rho+B(t_n)}}=\frac{1}{I(\rho)}. \quad (A7)$$

3) The conjugate prior. If the prior can provide the spin information, it can be set as a conjugate, which is the Gaussian distribution, so as to simplify the derivation in order to bring the posterior and the prior to the same distribution family[44]. In that regard, by considering the conjugate prior as $(\rho|f_0^{meas})\sim N(\rho_0,\sigma_0^2)$, when it is an unbiased estimate, Eq. (A1) can be simplified as follows:

$$-\sum_{n=1}^{N}\frac{A(t_n)}{(A(t_n)\cdot\rho+B(t_n))}(A(t_n)\cdot\rho^E+B(t_n)-f_n)-\frac{1}{\sigma_0^2}(\rho-\rho_0)=0. \quad (A8)$$

After carrying out the simplification procedure, the estimated results can be obtained:

$$\rho^E = \frac{\sum_{n=1}^{N}\frac{A(t_n)}{(A(t_n)\cdot\rho+B(t_n))}(f_n-B(t_n))+\frac{\rho_0}{\sigma_0^2}}{\sum_{n=1}^{N}\frac{A(t_n)^2}{(A(t_n)\cdot\rho+B(t_n))}+\frac{1}{\sigma_0^2}}. \tag{A9}$$

The variance $\sigma_{conjugate}^2$ of Jeffreys prior can be obtained by substituting the Poisson variance of the fluorescence signal $f_n$:

$$\sigma_{conjugate}^2 = 1/\left(\sum_{n=1}^{N}\frac{A(t_n)^2}{A(t_n)\cdot\rho+B(t_n)}+\frac{2}{\sigma_0^2}+1/\left(\sigma_0^4\sum_{n=1}^{N}\frac{A(t_n)^2}{A(t_n)\cdot\rho+B(t_n)}\right)\right) = \frac{1}{I(\rho)+\frac{2}{\sigma_0^2}+\frac{1}{\sigma_0^4\cdot I(\rho)}}. \tag{A10}$$

*Appendix B. Bayesian estimation of the system-on-chip architecture*

The system-on-chip architecture is a ZYNQ-based hardware platform, which mainly includes a ZYNQ board, a data acquisition (DAQ) board, and a pulse board. The overall design of the hardware is schematically illustrated in Fig. A1.

The ZYNQ board is the ZCU102 evaluation board from the Xilinx Company, which is equipped with a 4 GB fourth-generation DDR, a clock, an Ethernet cable, a USB port, and other chips. It also includes an FPGA part and an ARM part, in which the FPGA logic contains high-speed data acquisition and pulse generation units. For the data acquisition, the JESD204B module converts high-speed serial data from AD9680 to parallel, and the clock recovery technology is used to recover the 312.5 MHz clock applied for the parallel data processing. A Trim signal interception module receives parallel data from JESD204B and detects pulse trigger signals. After a trigger signal has been detected, the module starts to store data of specified length to the BRAM and generates an interrupt signal to the ARM at the same time. For the pulse generator part, a 312.5 MHz clock is employed to generate a 500 MHz clock through the phase-locked loop, which outputs a pulse with a 2 ns resolution. The ARM uses the AXI-Lite bus interface for the parameter configuration and operation control. At the same time, the ARM realizes both data receipt and real-time Bayesian estimation under the trigger of the interrupt. The data interaction of the whole system is realized by means of LabVIEW software, including the instruction of measurement parameters and the receipt of the readout results during the Bayesian estimation.

The data acquisition and pulse boards are connected to the ZCU102 board through an FPGA Mezzanine Card (FMC) interface with high-density pins and high bandwidth characteristics. After performing a single-end differential conversion, the common-mode voltage bias and the voltage gain adjustment procedures are realized by employing a signal conditioning circuit consisting of low-noise amplifier chips ADA4932 and LMH6401. The fluorescence measurement signal is then transmitted to the data acquisition sub-board. Driven by the clock chip AD9528, the data acquisition sub-board with the AD9680 converter as the core digital-to-analog conversion unit digitizes analog signals and transmits them to the ZCU102 board. The ZYNQ chip of the ZCU102 board afterward generates the original pulse signal and transmits it to the pulse board through the FMC interface. The transistor-transistor logic pulse output board possesses three output channels, where each channel uses SN74AVC1T45 and BUF602 to adjust the voltage of the original pulse signal to the desired level.

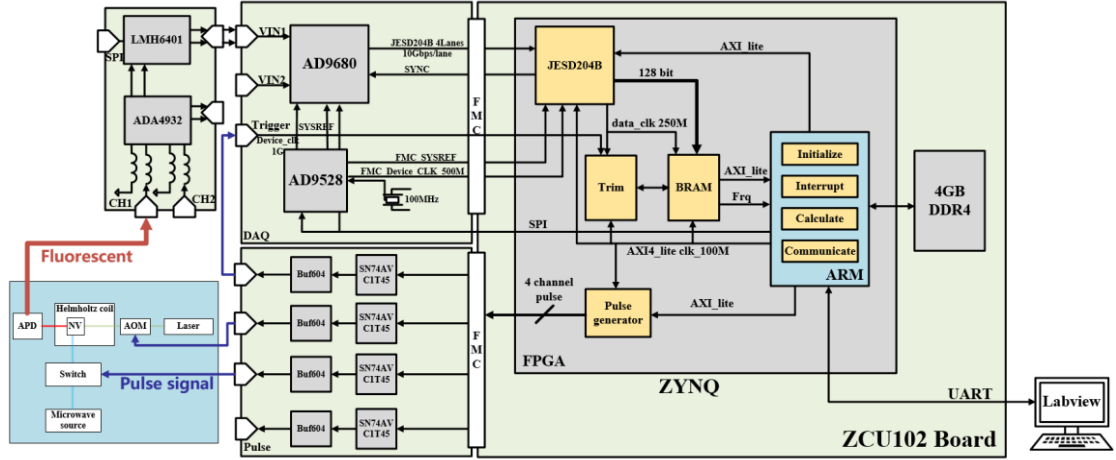

Figure A1. Architecture of the ZYNQ-based hardware platform

*Appendix C. Parameters estimation of the likelihood function*

Both the parameters $\lambda$ and $L$ in the likelihood function need to be determined according to the actual situation. 1) The parameter $\lambda$ is only used as the coefficient term of the likelihood function signal, so the ratio of the theoretical and actual signals in the steady-state can be employed as the estimation of this parameter. 2) The maximum optical pumping rate $L \cdot r$ corresponds to the maximum laser power was measured to be ~ 40 MHz (corresponding to $L \approx 0.9$). The laser noise is the pink noise, and the time of one measurement sequence is on the order of a microsecond, so the parameter $L$ can be considered constant during this period. The obtained estimation result of $\rho^E \approx 1$ without microwave, and the estimation result $\rho^E \gg 0$ after $\pi$ microwave due to the influence of relaxation and pulse error. The estimation method of the parameter $L$ is illustrated in Fig. A2. After collecting the reference signal in the sequence and calculating $\lambda$ by the steady-state value, a larger parameter $L$ is firstly selected to indued the signal included in the theoretical signal. The theoretical curve at $\rho = 1$ approaches the experimental value by decreasing the $L$ value gradually. It should be noted that since the signal containing noise may exceed the theoretical curve at $\rho = 1$, a certain margin should be left to contain noise.

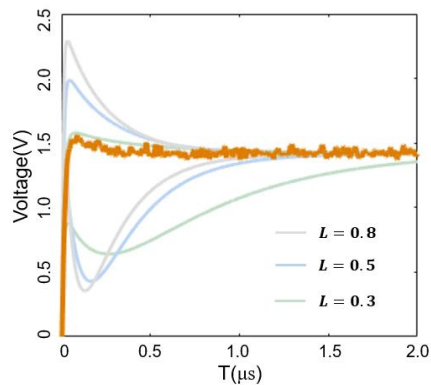

Figure A2. Example of the parameter estimation during the Bayesian update. The yellow curve is the experimental signal of the APD collected by ADC, and the light-colored curve is the theoretical fluorescence curve of $\rho = 1$ and $\rho = 0$ by employing different $L$ conditions.